\documentclass[runningheads]{llncs}
\usepackage[T1]{fontenc}
\usepackage{svg}
\usepackage{amsmath}
\usepackage[hidelinks]{hyperref}
\usepackage{makecell, cellspace, caption}
\usepackage{float}
\usepackage{array}
\usepackage{booktabs}
\newcolumntype{H}{>{\setbox0=\hbox\bgroup}c<{\egroup}@{}}

\svgpath{{../images/}} 
\usepackage{graphicx}

\usepackage{color}

\begin{document}
\title{Network Entropy as a Measure of Socioeconomic Segregation in Residential and Employment Landscapes}
\titlerunning{Entropy as a Measure of Socioeconomic Segregation}
\author{
\href{https://orcid.org/0000-0002-5683-3023} {\includegraphics[scale=0.06]{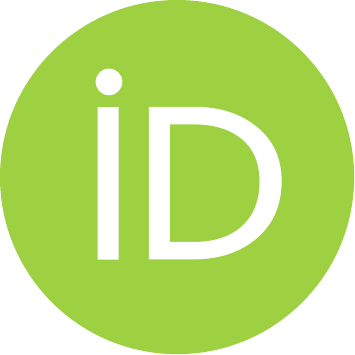}\hspace{1mm}Nandini Iyer}\inst{1}
\and
\href{https://orcid.org/0000-0002-6479-6429} {\includegraphics[scale=0.06]{images/orcid.pdf}\hspace{1mm}Ronaldo Menezes}\inst{1,2}
\and 
\href{https://orcid.org/0000-0002-3927-969X} {\includegraphics[scale=0.06]{images/orcid.pdf}\hspace{1mm}Hugo Barbosa}\inst{1}}
\authorrunning{N. Iyer et al.}
\titlerunning{Network Entropy as a Measure of Socioeconomic Segregation}
\institute{BioComplex Laboratory, Computer Science, University of Exeter, UK \and
  Computer Science, Federal University of Cear\'a, Fortaleza, Brazil
  \\
  \email{$^{*}$niyer@biocomplex.org, r.menezes@exeter.ac.uk, h.barbosa@exeter.ac.uk}}%\and
\maketitle              % typeset the header of the contribution

\begin{abstract}
Cities create potential for individuals from different backgrounds to interact with one another. It is often the case, however, that urban infrastructure obfuscates this potential, creating dense pockets of affluence and poverty throughout a region. The spatial distribution of job opportunities, and how it intersects with the residential landscape, is one of many such obstacles. In this paper, we apply global and local measures of entropy to the commuting networks of 25 US cities to capture structural diversity in residential and work patterns. We identify significant relationships between the heterogeneity of commuting origins and destinations with levels of employment and residential segregation, respectively. Finally, by comparing the local entropy values of low and high-income networks, we highlight how disparities in entropy are indicative of both employment segregation and residential inhomogeneities. Ultimately, this work motivates the application of network entropy to understand segregation not just from a residential perspective, but an experiential one as well.

\end{abstract}

\section{Introduction}
        
Often, the prosperity of urban areas is understood as a function of economic input and output \cite{oecd2013compendium}. However, experiential variables such as social inclusion have been shown to fuel the productivity of cities \cite{diaz2021segregation,veneri2021divided}. Accordingly, investing in improving the diversity of cities can lead to positive socioeconomic outcomes, fostering innovation and entrepreneurship \cite{qian2013diversity}. The urban experience, in terms of mobility, is largely comprised of trips to work \cite{jiang2016timegeo}, underscoring the importance of considering social inclusion not only from the residential dimension, but the employment perspective as well. In this sense, mobility, specifically commuting patterns, can be viewed as a potential way to improve diversity and integration by creating points of social inclusion in employment areas \cite{hackl2018mobility}.

Networks are particularly useful for analysing commuting behaviours, as they capture structural patterns that other approaches may overlook \cite{louail2015uncovering}. Specifically, network entropy can capture the concentration of labour supply and demand as well as the level of diversity of where workers are commuting to or from \cite{marin2022uncovering}. Entropy has been used in commuting networks to explain economic growth \cite{goetz2010us}, identify spatial inequalities \cite{lenormand2020entropy}, and measure social assortativity \cite{bokanyi2021universal}. However, the majority of previous research on this topic
%% RMComment: "this" research refers to your paper, or to the research you have just mentioned above? It's a bit ambiguous. 
analyses the commuting networks of an entire population. Thus, we consider not only the commuting networks of an entire population, but also commuting networks comprised of individuals from particular socioeconomic groups. Disaggregating commuting networks by demographics allows us to study whether disparities in structural diversity could serve as an indicator of social exclusion.

In this paper, we explore the intersection of socioeconomic segregation in cities and structural diversity in commuting patterns by applying an information-theoretic approach to mobility networks. Specifically, we leverage measures of global and local entropy of commuting networks to clarify whether demographic disparities in reliance on commuting origins and destinations correspond with levels of segregation on a residential and employment level. Our analyses of the commuting networks of 25 US cities indicate that, for high-income workers, higher levels of residential concentration correlate with higher employment segregation. Employment segregation strays from residential segregation, in that it measures the level of segregation based on the individuals that work in a region, rather than individuals that live there.
%% RMComment: I'm not so sure people will understand what employment segregation means? Has this been defined?
Low-income workers, on the other hand, tend to commute to a larger range of workplace areas. Finally, our network approach to segregation reveals facets of structural socioeconomic inhomogeneities in commuting patterns beyond what traditional measures of segregation can capture.

\section{Methods and Data}
\subsection{Data}

We use the LEHD Origin-Destination Employment Statistics (LODES) dataset
%% RMComment: What does the L in LODES stand for? 
from the United States Census's 2019 Longitudinal Employer-Household Dynamics (LEHD) program \cite{us2020lehd}. The dataset captures the residential patterns of the surveyed workforce by measuring the number of individuals commuting from one census block group to another. We evaluate residential-employment trends on a census tract level. Census tracts are statistical partitions of counties that contain anywhere from 1,200 to 8,000 residents. We use the LODES data to construct commuting networks for 25 cities in the United States, where every node reflects a census tract and directed, weighted edges depict the number of individuals commuting from one tract to another. The 25 cities we selected cover a wide range of population sizes and socioeconomic characteristics. The LODES dataset also provides information about how the total commutes from a pair of census tracts are distributed across lower, middle, and higher-income demographics. The low-income group consists of individuals earning less than \$1,250 per month, while the minimum monthly income for the high-income group is \$3,333. In addition to building the entire commuting network of a city, we also build a low, middle, and high-income network, which have the same nodes as the network for the entire city. However, the networks for each socioeconomic group, which we refer to as disaggregated networks, have different edge weights depending on the number of people in a socioeconomic group that commute between a pair of tracts.

\begin{figure}
\centering
\includegraphics[width=\textwidth]{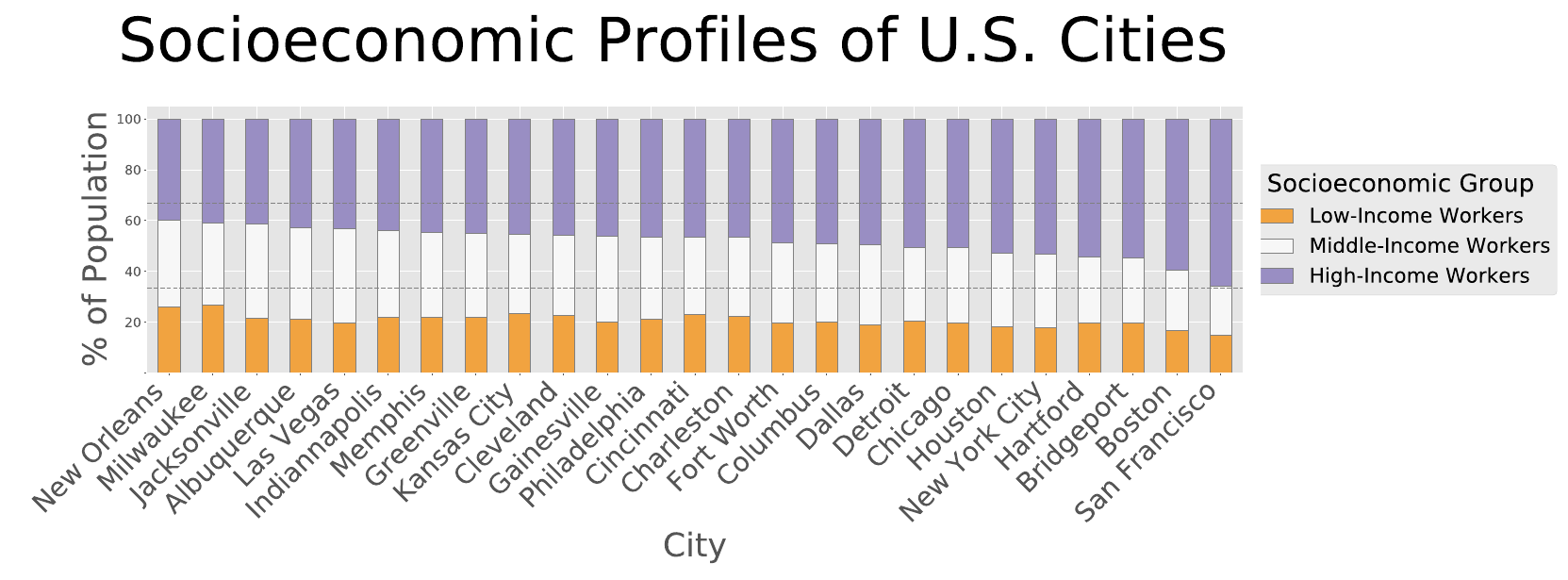}
\caption{Workforce distribution by income level split into three socioeconomic groups (low, middle and high-income) across 25 US cities. } \label{pop_distr}
\end{figure}

Figure \ref{pop_distr} captures the socioeconomic makeup of each city, according to the LODES data. The dashed lines indicate what an even distribution across demographics would look like. Thus, we observe that, within the context of the LODES dataset, %cities such as New Orleans and Milwaukee have more equal representation of socioeconomic groups than cities such as Boston and San Francisco. 
cities such as Boston and San Francisco have skewed representation of socioeconomic groups. To measure levels of residential segregation, we use Table B19001 from the 2019 American Community Survey 5-Year Estimates, which measures the population distribution across income brackets for each census tract \cite{us2020acs}.

\subsection{Network Entropy in Commuting Networks}

Throughout our analyses, we use the term \textit{commuting destinations} to refer to the workplaces that a residential population commutes to, while \textit{commuting origins} describe the residential areas from which an employment area's workforce commutes. Furthermore, \textit{labour supply} refers to employment areas that supply jobs, while \textit{labour demand} reflects the employable population of a region.  We use Shannon's entropy, which captures the level of information that can be extracted given a probability distribution, as a measure of diversity in commuting destinations and origins \cite{shannon1948mathematical}. Moreover, network entropy can be applied at different network resolutions and can focus on the commuting in-flow to a work area or the commuting out-flow from a residential region. Global entropy of in-flow and out-flow captures the urban concentration of labour supply and demand, respectively, by characterising the degree of monocentricity. That is, when a city has one area that supplies most of the labour opportunities, there is less uncertainty in predicting commuting destinations, corresponding with a lower global in-flow entropy value for the entire city. Thus, global in-flow entropy ($H_{GN}^{in}$) leverages the node strength, $\sum_{i}p_{ij}$, of all incoming commutes to each tract, $j$ in a city:

\begin{equation}
H_{GN}^{in} = \frac{-\sum_{\forall j}\left(\sum_{\forall i} p_{ij}\right)log\left(\sum_{\forall i} p_{ij}\right)}{log(n)},
\label{global_eq}
\end{equation}

\noindent where $n$ is the total number of tracts in a city. Global out-flow entropy captures the distribution of labour demand, such that low entropy values depict a scenario in which most workers live in a few areas and high entropy values indicate where residential locations of workers are more evenly distributed across nodes. In contrast to $H_{GN}^{in}$, global out-flow entropy ($H_{GN}^{out}$) calculates the out-degree node strength, $\sum_{j}p_{ij}$, for all census tracts, $i$ in a city.

Global entropy defines a city with respect to how labour supply or demand is distributed across all census tracts. Meanwhile, local entropy defines each tract based on how evenly distributed all its incoming or outgoing commutes are. A high local in-entropy for a census tract implies that the commuting origins for individuals who work in that tract are evenly distributed across all the potential origins.  Local in-flow entropy ($H_{L}^{in}$) for a census tract $j$ accounts for the probability, $p_{(i|j)}$, of tract $j$ receiving commutes from a census tract $i$, for all possible commuting origins, $i$:

\begin{equation}
H_{L}^{in} = \frac{-\sum_{\forall i}\frac{p_{ij}}{p_j}log\frac{p_{ij}}{p_j}}{log(n-1)}.
\label{local_eq}
\end{equation}

Local out-flow entropy ($H_{L}^{out}$) can be defined similarly, except rather than consider the probability $p_{(i|j)}$ for incoming commutes, it calculates the probability, $p_{(j|i)}$ of tract $i$ sending commuters to a census tract $j$, for all possible workplaces $j$. In this manner, we leverage network entropy measures on a global and local scale to understand how commuting characteristics of low versus high income workers indicate levels of segregation not only in terms of where each demographic group lives, but also in the context of the areas where they work. The denominators in Equations \ref{global_eq} and \ref{local_eq} serve to normalise the entropy values for cities of varying sizes, based on the number of tracts in a city (Eq. \ref{global_eq}), and the focal node's maximal possible degree (Eq. \ref{local_eq}).

\subsection{Characterising Segregation in Urban Landscapes}

We analyse features of the urban space to better understand the mechanisms that drive the identified differences in commuting networks across demographics. For such, we characterise urban areas in terms of the concentration of socioeconomic groups, using the Index of Concentration at the Extremes (ICE), proposed by Douglas Massey in an attempt to describe a region using both spatial attributes and the characteristics of polarised demographics \cite{booth2001prodigal}. While most segregation metrics account for how segregated the minority group is in relation to the entire population, ICE incorporates both ends of the demographic spectrum:

\begin{equation}
\text{ICE}_{i} = \frac{A_i - P_i}{T_i},
\label{ICE}
\end{equation}

\noindent where the ICE for a census tract $i$ is defined as the difference between the number of affluent residents, $A_i$, and the residents under the poverty line, $P_i$, over the entire population, $T_i$. While the numerator captures the imbalance between the extremes, the denominator expresses the degree of imbalance in relation to the entire population of tract $i$.  Thus, ICE aims to measure the imbalance between affluence and poverty by measuring the concentration of both the extremely disadvantaged and advantaged in a given population. In a similar vein, we can use Equation \ref{ICE} to measure the level of segregation for a census tract, $i$, from an employment perspective, defining $A_i$ and $P_i$ as the number of workers commuting to tract $i$ from the high and low-income group, respectively. Thus, $T_i$ captures the total number of individuals commuting to $i$, regardless of demographics. Accordingly, we can use ICE to capture segregation levels at both a residential ($\text{ICE}_{\text{res}}$) and employment ($\text{ICE}_{\text{emp}}$) scale. We use the LODES dataset to calculate $\text{ICE}_{\text{emp}}$ as it provides unique information regarding demographic characteristics of a workforce. Meanwhile, We use the ACS median household income distributions to calculate residential segregation as it provides a high granularity of income-levels to characterise tracts by its residents.

\section{Results}

\subsection{Global Entropy and City-Level Analyses}

\begin{figure}
\centering
\includegraphics[width=\textwidth]{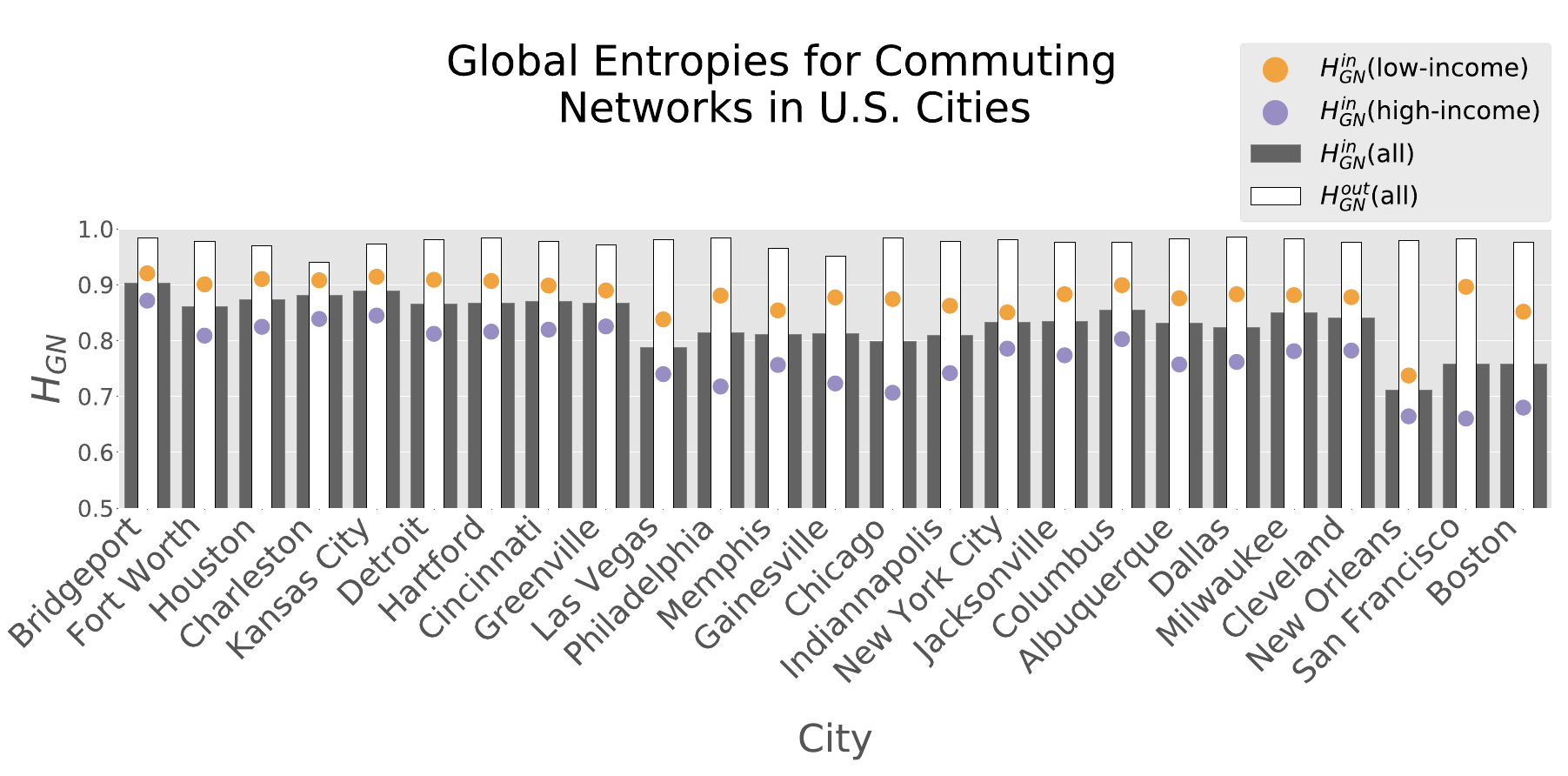}
\caption{Global entropy values for 25 US cites. The grey and white bar capture global entropy of the entire commuting network for in-flow and out-flow commutes, respectively. The orange point measures $H_{GN}^{in}$ for the low-income network, while the purple point does the same for the high income network.} \label{global_entropy}
\end{figure}

We begin by applying global entropy measures to the commuting networks of the 25 different cities. We reiterate that because entropy values are normalised with respect to network size, we can make comparisons not only between cities, but also between commuting networks of different socioeconomic groups. Figure \ref{global_entropy} elucidates how, for every city, the global entropy of commuting out-flows is consistently larger than the global entropy for commuting in-flows. This pattern is to be expected, as residential locations are known to be more evenly distributed than employment hubs \cite{marin2022uncovering}. Notably, the lower values of global in-flow entropy in cities such as New Orleans, San Francisco, and Boston indicate the presence of larger employment hubs.

By disaggregating the commuting networks based on workers' economic profiles, we can analyse the networks of low-income and high-income workers separately. Interestingly, within the cities we analyse, the global in-flow entropy of high-income commuters is persistently lower than that of the low-income group. These lower values imply less structural diversity in high-income commuting origins, which translates to a higher level of monocentricity for high-income jobs. What is clear is that there is a distinct difference in global entropy values for commuting in-flows when considering networks of different socioeconomic groups. Whether these differences are indicative of socioeconomic inequality is explored in the following sections. 

\subsection{Local Entropy as a Measure of Segregation}
%% RMComment: I'm not sure the section names, or subsections should be that long. Generally titles are short. 

In this section, we aim to disentangle whether overall structural diversity entails other forms of diversity, such as lower levels of segregation. In doing so, our goal is to clarify if the monocentricity of job opportunities, which is more present in high-income networks, is a privilege or a burden. 

\subsubsection{Residential segregation}
To better understand the differences in how housing and employment landscapes intersect for socioeconomic groups, we evaluate whether a relationship exists between residential segregation and the diversity of employment destinations for residents of a particular tract. %The first few columns in Table \ref{local_entropy_corr} list general network properties of each city's entire commutng netowrk. 
The fourth column of Table \ref{local_entropy_corr} lists the Pearson correlation coefficients when comparing the local out-flow entropy ($H_{L}^{\text{out}}$) with the residential ICE value (Eq. \ref{ICE}) of census tracts in a city. We recall that the higher the $\text{ICE}_{\text{res}}$ value, the larger the proportion of high-income residents in that area.  All but three of the 25 cities have a significant, negative correlation.  This reveals that census tracts characterised by more affluent residents tend to have lower local diversity values (i.e., concentrate commutes to fewer tracts), which indicate their dependence on particular tracts for supplying labour to their residents.  On one hand, one can argue that these higher diversity values for less affluent tracts makes them less vulnerable to any shortages in labour supply, such that if an employment location stops providing opportunities, they have other options of commuting destinations. However, this negative correlation could also indicate the presence of inequalities in the housing landscape. Thus, we proceed to analyse the opposite dynamic, comparing the diversity of commuting origins ($H_{L}^{in}$) to the degree of employment segregation ($\text{ICE}_{\text{emp}}$), to explore whether the identified negative correlations imply socioeconomic disparity on a residential or employment level.

\subsubsection{Employment segregation}
For completeness, we have included the results for the correlation patterns between segregation of a tract's workforce ($\text{ICE}_{\text{emp}}$, Equation \ref{ICE}) and the diversity of commuting origins for that workforce ($H_{L}^{in}$) in the fifth column of Table \ref{local_entropy_corr}, which shows significant positive correlations for 20 of the cities.  When considering the high-income group, we see that, from an employment perspective, tracts in which more affluent employees work are more diverse, expressing heterogeneity in commuting origins. However, from a residential perspective, tracts with more affluent residents are less diverse, expressing homogeneity in commuting destinations. Thus, we show that higher values of structural diversity do not necessarily imply an advantage.

\begin{table}[ht!]
\caption{General network properties and Pearson correlation coefficients for different forms of local entropy and socioeconomic segregation. The last column refers to local entropy differences between the disaggregated networks, discussed in Section \ref{ssec:employment_seg}. Boldface is used to indicate significant correlations, with asterisks reflecting p-value.}\label{tab1}
\centering
\renewcommand{\arraystretch}{1.0}
\begin{tabular}{lp{1.7cm}Hp{1.7cm}Hp{1.7cm}Hp{1.7cm}Hp{1.7cm}H}
\toprule
\multicolumn{1}{c}{} & \multicolumn{4}{c}{Network} & \multicolumn{2}{c}{$H_{L}^{out}$ vs. } & \multicolumn{2}{c}{$H_{L}^{in}$ vs.}  & \multicolumn{2}{c}{$\Delta H_{L}^{in}$ vs.}\\

\multicolumn{1}{c}{City} & \multicolumn{4}{c}{Properties} & \multicolumn{2}{c}{$ICE_{res}$} & \multicolumn{2}{c}{$ICE_{emp}$} & \multicolumn{2}{c}{$ICE_{emp}$} \\
\cmidrule{2-10} 
& \textit{Nodes} & \textit{p} & \textit{Edges} & \textit{p}  & \textit{$r^1$} & \textit{p} &  \textit{$r^1$} &  \textit{p} &  \textit{$r^1$}  &  \textit{p}  \\
\midrule

Charleston & 85 & & 6,162 & & \textbf{-0.313**} & 0.004 & 0.116 & 0.292  & \textbf{0.453***} & 0.000\\

San Francisco & 196 & & 25,886 & & \textbf{-0.373***} & 0.000 & \textbf{0.614***} & 0.000 & \textbf{0.341***} & 0.000 \\

Gainesville  & 56 & & 2,803 & & -0.243 & 0.072 & \textbf{0.313*} & 0.019  & \textbf{0.496***} & 0.000\\

Greenville  & 111 & & 10,431 & & \textbf{-0.386***} & 0.000 & 0.033 & 0.735  & \textbf{0.596***} &  0.000\\

Albuquerque & 153 & & 18,871 & &  \textbf{-0.385***} & 0.000 & 0.116 & 0.154 & \textbf{0.632***} &  0.000\\

New Orleans & 176 & & 13,680 & & \textbf{-0.309***} & 0.000 & 0.140 & 0.068  & \textbf{0.843***} &  0.000\\

Houston & 921 & & 398,876 & & \textbf{-0.287***} & 0.001 & -0.018 & 0.832 & \textbf{0.716***} &  0.000\\

Boston & 204 & & 20,608 & & \textbf{-0.401***} & 0.000 & \textbf{0.652***} & 0.000  & \textbf{0.646***} &  0.000 \\

Indianapolis & 224 & & 33,147 & & \textbf{-0.494***} & 0.000 & \textbf{0.231***} & 0.001  & \textbf{0.754***} &  0.000\\

Las Vegas & 487 & & 124,083 & &  0.027 & 0.551 & \textbf{0.093*} & 0.039 & \textbf{0.818***} & 0.000\\

Philadelphia  & 384 & & 69,364 & & \textbf{-0.547***} & 0.000 & \textbf{0.289***} & 0.000  & \textbf{0.754***} &  0.000 \\

Columbus & 347 & & 76,995 & & \textbf{-0.486***} & 0.000 & \textbf{0.272***} & 0.000  & \textbf{0.735***} &  0.000\\ 

Hartford  & 224 & & 33,107 & &  \textbf{-0.507***} & 0.000 & \textbf{0.430***} & 0.000  & \textbf{0.710***} &  0.000\\

Jacksonville & 173 & & 23,610 & & \textbf{-0.441***} & 0.000 & \textbf{0.473***} & 0.000  & \textbf{0.722***} & 0.000\\

Cincinnati & 222 & & 32,801 & & \textbf{-0.185**} & 0.006 & \textbf{0.306***} & 0.000 & \textbf{0.751***} &  0.000\\

Milwaukee  & 297 & & 50,210 & &  \textbf{-0.517***} & 0.000 & \textbf{0.377***} & 0.000  & \textbf{0.829***} & 0.000\\

Cleveland & 446 & & 85,609 & & \textbf{-0.182***} & 0.000 & \textbf{0.330***} & 0.000 & \textbf{0.834***} &  0.000 \\

Bridgeport & 210 & & 28,259 & &   \textbf{-0.425***} & 0.000 & \textbf{0.278***} & 0.000  & \textbf{0.665***} &  0.000 \\

Fort Worth & 357 & & 76,883 & & \textbf{-0.483***} & 0.000 & \textbf{0.280***} & 0.000  & \textbf{0.798***} &  0.000\\

Memphis & 221 & & 31,507 & & \textbf{-0.135*} & 0.047 & \textbf{0.212**} & 0.002 & \textbf{0.816***} & 0.000 \\

Chicago & 1,318 & & 441,406 & &  \textbf{-0.287***}& 0.000 & \textbf{0.358***} & 0.000  & \textbf{0.855***} &  0.000\\

New York City & 2,164 & & 990,302 & &  \textbf{-0.478***} & 0.000 & \textbf{0.455***} & 0.000  & \textbf{0.837***} & 0.000 \\

Detroit & 1,163 & & 385,185 & & \textbf{0.139***} & 0.000 & \textbf{0.527***} & 0.000 & \textbf{0.865***} &  0.000\\

Dallas & 529 & & 129,486 & & \textbf{-0.569***} & 0.000 & \textbf{0.350***} & 0.000  & \textbf{0.830***} &  0.000 \\

Kansas City & 283 & & 49,377 & &  \textbf{-0.165**} & 0.006 & \textbf{0.328***} & 0.000 & \textbf{0.781***} & 0.000\\
\bottomrule
\multicolumn{9}{l}{$^1 \text{*p<0.05; **p<0.01; ***p<0.001}$} \\

\end{tabular}
\label{local_entropy_corr}
\end{table}

This section elucidated how the structural diversity of census tracts often corresponds with their urban characteristics. We measure the local in and out-flow entropy of entire commuting networks to highlight how diversity of the employment landscape can be reflective of employment and residential segregation. The next section wraps up this analysis by disaggregating the entire commuting network of each city into low and high-income networks. This allows us to examine how unequal labour distribution may be exacerbating existing inequalities.

\subsection{Socioeconomic Disparities in Diversity of Commuting Origins}\label{ssec:employment_seg}

In this section, we illustrate how local in-flow entropy measures of disaggregated networks can capture experienced segregation. For improved readability and due to space constraints, we show the results for four representative cities of different sizes and socioeconomic profiles (Milwaukee, San Francisco, New York City and Detroit). Nevertheless, our findings are based on the analyses of the 25 cities.

\begin{figure}[htb]
\centering
\includegraphics[width=1\textwidth]{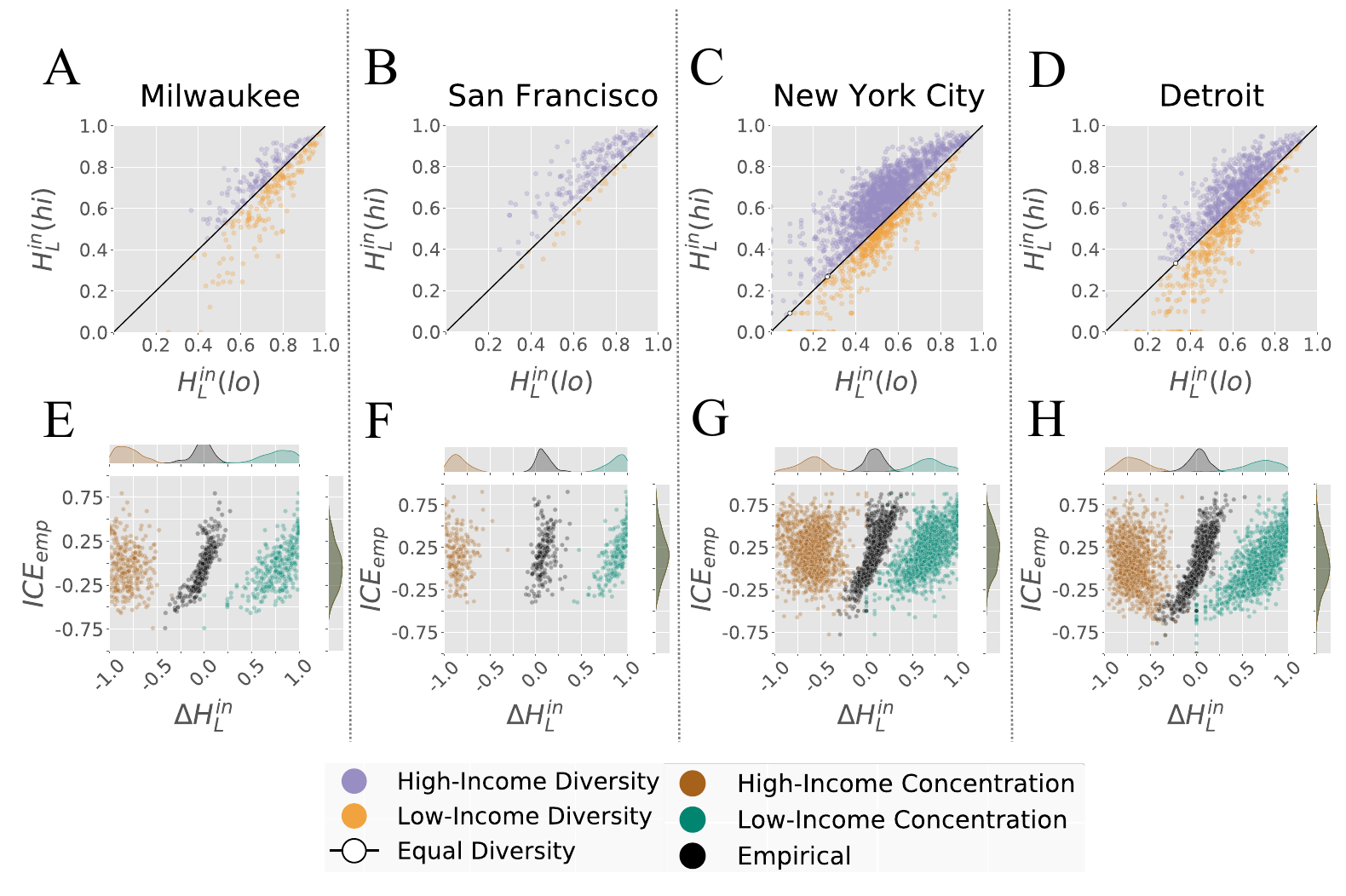}
\caption{Local in-flow entropies for high and low-income networks in 4 US cities. Panels A-D compare entropy values for tracts in the low and high-income network, with points below the diagonal reflecting tracts in which low income workers have more diverse commuting origins. Panels E-H show how the differences in these values (black points) compare to null models derived from Figure \ref{toy_example}} \label{disaggregated_res}
\end{figure}

We proceed, analysing how differences in heterogeneity of commuting origins coincide with levels of employment segregation. We begin by splitting the commuting network of the city into separate networks that measure the residential-work patterns of socioeconomic groups separately. In doing so, we can understand the extent to which the diversity of commuting origins differs between the low-income and high-income workers in a census tract. Panels A-D in Figure \ref{disaggregated_res} plot the local in-flow entropies of every tract in the low-income network ($H_{L, \textit{lo}}^{in}$) against their respective entropy values in the high-income network ($H_{L, \textit{hi}}^{in}$), for four of the analysed cities. The black line expresses the case in which a census tract has equal diversity of commuting origins in both socioeconomic networks, which we see a few cases of in New York City and Detroit, indicated by the white points on the diagonal. Orange points reflect census tracts in which the low-income individuals have a more even distribution of commuting origins than the high-income workforce. The purple points capture census tracts with the opposite characteristics: greater diversity in residential locations for the affluent workforce. We observe that most census tracts in San Francisco and New York City tend to have higher homogeneity in commuting origins for low-income workers than compared to high-income workers. This comparison opens the gateway into using local entropy values to extend our understanding of segregation from a residential dimension to an employment one as well.

In order to accomplish this, we evaluate how socioeconomic disparities in the homogeneity of residential locations for a region relate to the demographic balance of that region's workforce.  For each tract in a city, we compare a census tract's in-flow entropy value in the high and low-income networks:

\begin{equation}
\Delta H_{L}^{in}(i) =  H_{L, \textit{hi}}^{in}(i) - H_{L, \textit{lo}}^{in}(i),
\label{local_diff}
\end{equation}

\noindent where $H_{L, \textit{hi}}^{in}(i)$ captures the local in-flow entropy of census tract $i$ in the high-income commuting network, whereas $H_{L, \textit{lo}}^{in}(i)$ describes the in-flow entropy for $i$ in the low-income commuting network. Thus, $\Delta H_{L}^{in}(i)$ can range from $-1$ to $1$, where negative values represent more heterogeneity of commuting origins for the low-income population. Positive values capture scenarios in which higher-income workers have more heterogeneous commuting origins than lower-income workers. 

\begin{figure}[htbp]
\centering
\includegraphics[width=\textwidth]{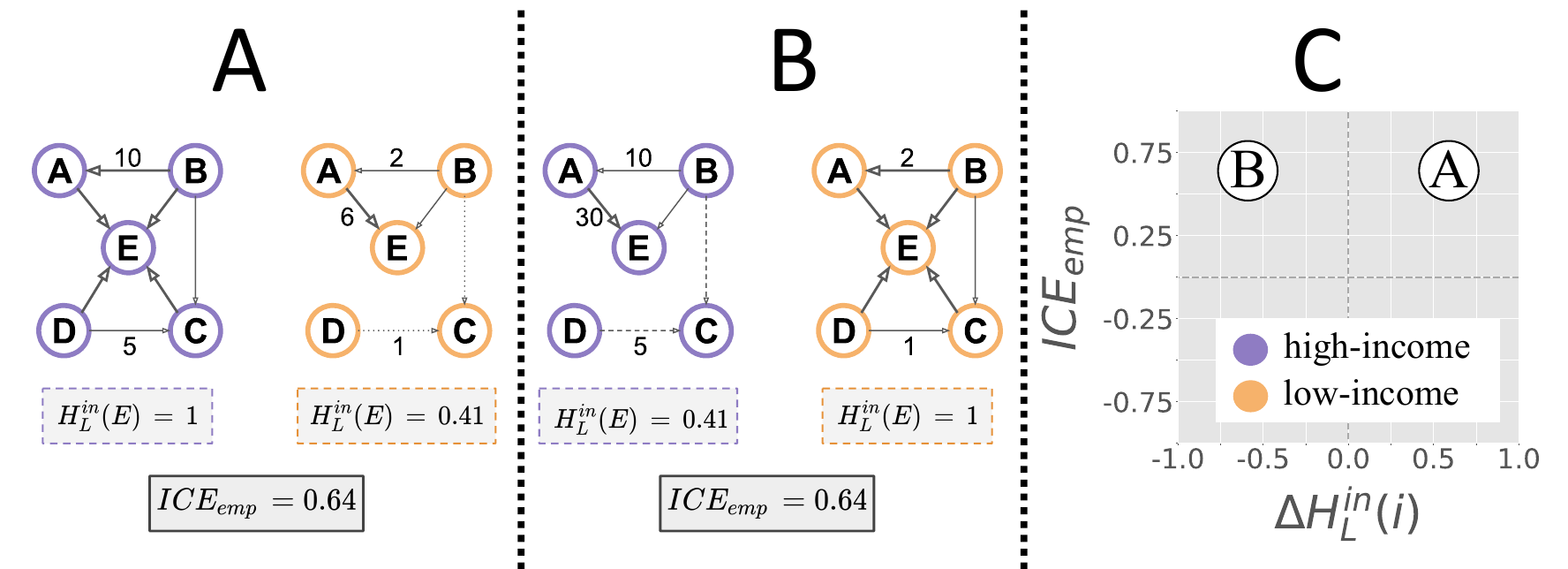}
\caption{Toy example highlighting the distinction between $\Delta H_{L}^{in}(i)$ and $ICE_{emp}$. Panel A shows concentration of high-incoming commuting origins, while B captures homogeneous origins for the low-income group. Panel C conveys how the two scenarios of commuting diversity relate to employment segregation.} \label{toy_example}
\end{figure}

\subsubsection{Comparing Local Entropies in Disaggregated Networks}
For each of the 25 cities, we find significant positive correlations between the difference in local entropy values ($\Delta H_{L}^{in}(i)$, Eq. \ref{local_diff}) to the level of segregation in employment areas ($\text{ICE}_{\text{emp}}$, Eq. \ref{ICE}). These correlations are outlined in the last column of Table \ref{tab1}.  If we consider the local in-flow entropy of tract $i$ in the high-income network, we can measure how evenly the total number of affluent individuals commuting to $i$ is distributed across all other census tracts in the city. We use Figure \ref{toy_example} to explain how entropy values are not strictly correlated with in-degrees.  The purple network captures the high-income commuting network and the orange network reflects that of the low-income workers. We set the total number of individuals working in node $E$ to be 50. For Scenarios A and B, we can observe that the socioeconomic composition of individuals working in node E remains the same (40 high-income workers, 8 low-income workers). Thus, values of employment segregation ($\text{ICE}_{\text{emp}}$) are consistent throughout the examples. What changes across the scenarios are the values of $H_{L}^{in}$, which describe how evenly commuting origins are distributed across other nodes in the network. Scenario A depicts a case where low-income commutes are more concentrated and high-income commuting origins are more diverse. Meanwhile, Scenario B captures homogeneity in high-income commuting origins and heterogeneous commuting origins for the low-income workers in the node $E$. We can, then, understand that $\Delta H_{L}^{in}(i)$ serves to measure differences in heterogeneity of commuting origins between the high and low income networks. The right most panel in Figure \ref{toy_example} uses Scenarios A and B to elucidate how positive correlations between $ICE_{emp}$ and $\Delta H_{L}^{in}(i)$ are non-trivial.

\subsubsection{Null Model Comparisons}
The black scatter plots in Figure \ref{disaggregated_res}E-H illustrate the aforementioned positive correlations between socioeconomic differences in the heterogeneity of residences ($\Delta H_{L}^{in}(i)$) and employment segregation $\text{ICE}_{\text{emp}}$, for four cities. We use null models, inspired from the examples in Scenario A and B of Figure \ref{toy_example}, to emphasise how the observed positive correlations are not a result of measuring related network attributes. The null models elucidate how the correlations between employment segregation and local entropy of networks are not always significant or positive. Both null models retain the in-degree of the disaggregated, empirical commuting networks, only changing how evenly a node's incoming edges are distributed across other nodes in the network. 

The Low-Income Concentration null model, represented by the blue scatter plot, hypothesises that the observed positive correlation between commuting origin diversity and employment segregation is a consequence of greater diversity of origins for the affluent population than the low-income group, captured by positive values of $\Delta H_{L}^{in}(i)$. We construct this null model, such that the residential locations for a tract's high-income workforce is uniformly sampled across all other tracts. The origins of the low-income workforce are defined by one, randomly sampled tract, producing smaller values of $H_{L, \textit{lo}}^{in}(i)$ All the while, we maintain the empirical workplace composition for each census tract, thus retaining empirical values of $\text{ICE}_{\text{emp}}$ and aligning the Low-Income Concentration model with Scenario A in the toy example. Meanwhile, the High-Income Concentration model, inspired by Scenario B in the toy example, and the brown scatter plot in Figure \ref{disaggregated_res}, capture negative values of $\Delta H_{L}^{in}(i)$ by simulating high diversity of low-income commuting origins and homogeneity for high-income origins. 

We emphasise that, in both null models, the only difference from the empirical commuting network is from the distribution of incoming edges to a tract.  The equivalent distributions for $\text{ICE}_{\text{emp}}$, shown along the y-axis, demonstrate how measures of employment segregation remain consistent across the empirical and null model scenarios, despite having disparities in residential-work patterns. On the other hand, the distributions along the x-axis, illustrate how $\Delta H_{L}^{in}(i)$ can capture these structural inequalities by identifying tracts in which socioeconomic groups express stark differences in their dependence on commuting origins.  While the empirical scatter plot does not show signs of extreme structural disparities, as expressed by the null models' scatter plots, the positive correlations indicate that areas which have more heterogeneity of residential locations for a particular socioeconomic group tend to be areas in which that socioeconomic group is more concentrated. %That is, larger absolute values of $\Delta H_{L}^{in}(i)$ coincide with more employment segregation. 
 The sign of $\Delta H_{L}^{in}(i)$ specifies which demographic is segregated, with values less than zero implying low-income employment segregation.  Thus, this section highlights how measuring disparities in structural diversity of disaggregated networks can expose dimensions of segregation in residential-workplace dynamics, that conventional metrics of segregation may overlook.

\section{Conclusion}

This work highlights how network entropy can be used to capture sociodemographic inequalities in commuting patterns that classical segregation metrics fail to detect. We compare global in-flow entropies of 25 commuting networks across the U.S. to identify cities, such as San Francisco and Boston, that have higher degrees of monocentricity. 

Then, by incorporating local entropy measures for the entire commuting network, we uncover that census tracts with a higher concentration of affluence have residents that travel to more homogenous workplaces. Meanwhile, tracts that attract a higher-income workforce express trends of heterogeneity in commuting origins. Finally, by splitting cities' commuting networks into high and low-income networks, we demonstrate the strength of network entropy in identifying disparities in resident-workplace trends across socioeconomic groups. We find that larger differences in node-level entropies, which measures socioeconomic disparities of a work force in terms of their residential distribution, correspond with higher levels of employment segregation.

This brings into question what has long been deliberated in sociological fields: how the consequences of segregation may change depending on which demographic is segregated. Future work can examine how these disparities in commuting flows reflect in other aspects of urban life. 
Specifically, other null models can be explored to understand what mechanisms may be fuelling the strong correlation between entropy and segregation. Moreover, this framework can be applied to other mobility networks in the context of various demographic dimensions, such as gender or age.

\bibliographystyle{splncs04}
\bibliography{references}

\begin{thebibliography}{10}
\providecommand{\url}[1]{\texttt{#1}}
\providecommand{\urlprefix}{URL }
\providecommand{\doi}[1]{https://doi.org/#1}

\bibitem{bokanyi2021universal}
Bok{\'a}nyi, E., Juh{\'a}sz, S., Karsai, M., Lengyel, B.: Universal patterns of
  long-distance commuting and social assortativity in cities. Scientific
  reports  \textbf{11}(1),  1--10 (2021)

\bibitem{booth2001prodigal}
Booth, A., Crouter, A.C.: The prodigal paradigm returns: ecology comes back to
  sociology. In: Does it take a village?, pp. 53--60. Psychology Press (2001)

\bibitem{us2020acs}
Bureau, U.C.: 2019 american community survey 5-year estimates, table b19001
  (2022), accessed on 15/12/2022

\bibitem{us2020lehd}
Bureau, U.C.: Lehd origin-destination employment statistics data (2002-2019)
  (2022), accessed on 15/12/2022, Longitudinal-Employer Household Dynamics
  Program, LODES 7.5

\bibitem{diaz2021segregation}
Diaz, R., Garrido, N., Vargas, M.: Segregation of high-skilled workers and the
  productivity of cities. Regional Science Policy \& Practice  \textbf{13}(5),
  1460--1478 (2021)

\bibitem{goetz2010us}
Goetz, S.J., Han, Y., Findeis, J.L., Brasier, K.J.: Us commuting networks and
  economic growth: Measurement and implications for spatial policy. Growth and
  Change  \textbf{41}(2),  276--302 (2010)

\bibitem{hackl2018mobility}
Hackl, A.: Mobility equity in a globalized world: Reducing inequalities in the
  sustainable development agenda. World development  \textbf{112},  150--162
  (2018)

\bibitem{jiang2016timegeo}
Jiang, S., Yang, Y., Gupta, S., Veneziano, D., Athavale, S., Gonz{\'a}lez,
  M.C.: The timegeo modeling framework for urban mobility without travel
  surveys. Proceedings of the National Academy of Sciences  \textbf{113}(37),
  E5370--E5378 (2016)

\bibitem{lenormand2020entropy}
Lenormand, M., Samaniego, H., Chaves, J.C., da~Fonseca~Vieira, V., da~Silva,
  M.A.H.B., Evsukoff, A.G.: Entropy as a measure of attractiveness and
  socioeconomic complexity in rio de janeiro metropolitan area. Entropy
  \textbf{22}(3), ~368 (2020)

\bibitem{louail2015uncovering}
Louail, T., Lenormand, M., Picornell, M., Garcia~Cantu, O., Herranz, R.,
  Frias-Martinez, E., Ramasco, J.J., Barthelemy, M.: Uncovering the spatial
  structure of mobility networks. Nature communications  \textbf{6}(1), ~1--8
  (2015)

\bibitem{marin2022uncovering}
Marin, V., Molinero, C., Arcaute, E.: Uncovering structural diversity in
  commuting networks: global and local entropy. Scientific Reports
  \textbf{12}(1),  1--13 (2022)

\bibitem{oecd2013compendium}
OECD: Compendium of productivity indicators 2013 (2013)

\bibitem{qian2013diversity}
Qian, H.: Diversity versus tolerance: the social drivers of innovation and
  entrepreneurship in us cities. Urban Studies  \textbf{50}(13),  2718--2735
  (2013)

\bibitem{shannon1948mathematical}
Shannon, C.E.: A mathematical theory of communication. The Bell system
  technical journal  \textbf{27}(3),  379--423 (1948)

\bibitem{veneri2021divided}
Veneri, P., Comandon, A., Garcia-L{\'o}pez, M.{\`A}., Daams, M.N.: What do
  divided cities have in common? an international comparison of income
  segregation. Journal of Regional Science  \textbf{61}(1),  162--188 (2021)

\end{thebibliography}

\end{document}